\documentclass[12pt,fleqn]{article}
\usepackage{epsfig}
\newcommand{\nn}{\nonumber}

\newcommand{\be}{\begin{equation}}
\newcommand{\ee}{\end{equation}}
\newcommand{\bea}{\begin{eqnarray}}
\newcommand{\eea}{\end{eqnarray}}

\newcommand{\PCPV}{
P_{CP-odd}}
\newcommand{\PCPC}{
P_{CP-even}}

\newcommand{\np}[1]{Nucl.  Phys.  {\bf #1}}
\newcommand{\pl}[1]{Phys.  Lett.  {\bf #1}}
\newcommand{\pr}[1]{Phys.  Rev.  {\bf #1}}
\newcommand{\prl}[1]{Phys.  Rev.  Lett. { \bf #1}}

\makeatletter
\begin{document}

\title{\vskip-2.5truecm{\hfill \baselineskip 14pt {{
\small  \\
\hfill MZ-TH/00-14 \\ 
\hfill April 00}}\vskip .9truecm}
 {\bf Tau neutrinos from muon storage rings }}

\vspace{5cm}

\author{Gabriela Barenboim\footnote{\tt 
gabriela@thep.physik.uni-mainz.de} 
\phantom{.}and Florian Scheck\footnote{\tt 
Scheck@dipmza.physik.uni-mainz.de}
 \\  \  \\
{\it  Institut f\H ur Physik - Theoretische 
Elementarteilchenphysik }\\
{\it Johannes Gutenberg-Universit\H at, D-55099 Mainz, 
Germany}
\\
}

\date{}
\maketitle
\vfill

\begin{abstract}
\baselineskip 20pt
Charged tau leptons emerging in a long baseline experiment
with a muon storage ring and a far-away detector will positively
establish neutrino oscillations. We study the conversion of $\nu_\mu$
($\overline{\nu}_\mu$) and of $\overline{\nu}_e$ ($\nu_e$) to
$\nu_\tau$ or $\overline{\nu}_\tau$ for neutrinos from a 20~GeV muon
storage ring, within the strong mixing scheme and on the basis of the
squared mass differences which are compatible with all reported
neutrino anomalies, including the LSND data. In contrast to other
solutions which ignore the Los Alamos anomaly, we find charged tau
production rates which should be measurable in a realistic set up. As
a consequence, determining the complete mass spectrum of neutrinos as
well as all three mixing angles seems within reach. Matter effects are
discussed thoroughly but are found to be small in this situation.
\end{abstract}
\vfill
\thispagestyle{empty}

\newpage
\pagestyle{plain}
\setcounter{page}{1}

\section{Introduction}

Neutrino beams from a muon storage ring provide an ideal tool for the
next round of experiments, aiming at establishing quantitatively
oscillations and, possibly, CP-violation in the leptonic sector
\cite{FLC,geer}. The main reason for this is that both composition
and spectrum of such beams are perfectly known and, in addition,  the
beam energy may 
be tuned. For instance, the neutrino beams from a storage ring such as 
the ones described in \cite{FLC,geer} would contain an equal number of 
muon neutrinos and electron antineutrinos, or, depending on the sign
of the parent muon, an equal number of muon antineutrinos and electron 
neutrinos, without any contamination from other neutrino species. 

The physics potential of a storage ring and a far-away detector, with
regard to oscillations, was studied, e.g., in \cite{belen,belen2}, and,
with regard to CP-violation, in \cite{rom,nosCP}. In this paper we
address the question of appearance of tau neutrinos in the processes
\begin{displaymath}
  \nu_\mu \longrightarrow \nu_\tau \qquad \mbox{and} \qquad 
  \nu_e \longrightarrow \nu_\tau \, ,
\end{displaymath}
due to oscillations. We calculate the number of produced $\tau^+$ and
$\tau^-$, taking into account matter effects on the neutrino beam on
its way from the source to the detector, within the strong mixing
scheme of three flavours only that we had proposed earlier
\cite{nos}. The latter assumption is receiving growing support by the
fact that new data from KAMIOKANDE seem to disfavour the existence of
a fourth, sterile, neutrino, while still showing the characteristic
modulation of events as a function of zenith angle. As a
consequence of the strong mixing which helps to populate strongly the
$\nu_\tau$ and $\overline{\nu}_\tau $ final channels from both the
$\overline{\nu}_e$ ($\nu_e$) and the $\nu_\mu$ ($\overline{\nu}_\mu$)
initial states, we find $\tau^\pm$ production rates one to two orders
of magnitude larger than within a model ignoring the LSND data.

In the next section we summarize some relevant formulae and collect the
parameters extracted earlier from the analysis of all neutrino
anomalies. We then describe and discuss matter effects in the Earth's
crust and calculate the event rates for the $\tau^\pm$ appearance
channels. The article ends with our results as well as some
conclusions. 


\section{Formulae and parameters}

In a scenario with three families of leptons, the mixing between 
three neutrino species is described by a conventional
Cabibbo-Kobayashi-Maskawa matrix $U$ relating flavour states to mass
eigenstates. If neutrinos are Dirac particles, $U$ has the form

\bea\label{lCKM}
U=\pmatrix{c_{12}c_{13} & s_{12} c_{13} & s_{13} e^{-i \delta} \cr
-s_{12}c_{23}  - c_{12}s_{23}s_{13}e^{i \delta} & 
c_{12}c_{23}- s_{12}s_{23}s_{13}e^{i \delta} & s_{23}c_{13} 
\cr
s_{12}s_{23}- c_{12}c_{23}s_{13}e^{i \delta}  & 
- c_{12}s_{23}e^{i \delta} - s_{12}c_{23}s_{13} 
&c_{23}c_{13}}
\eea
where $c_{ij}=\cos\theta_{ij}$ and  $s_{ij}=\sin\theta_{ij}$. 
In the case they are of Majorana type, extra phases appear but their 
effects on neutrino observables are of order $m_\nu /E_\nu$ \cite{maj} 
and are generally negligible.

The oscillation probabilities in vacuum take the form \cite{lindner}
\bea
P(\nu_i \rightarrow \nu_j)& =& 
\PCPC(\nu_{i}\rightarrow\nu_{j}) + \PCPV(\nu_{i}\rightarrow\nu_{j}) \\
P(\bar{\nu}_{i}\rightarrow\bar{\nu}_{j}) &=&
\PCPC(\nu_{i}\rightarrow\nu_{j}) - \PCPV(\nu_{i}\rightarrow\nu_{j}),
\eea
where 
\bea
\PCPC(\nu_{i}\rightarrow\nu_{j}) &=& \delta_{ij} -4\mbox{Re}J^{ji}_{12}
\sin^2\Delta_{12} -4\mbox{Re}J^{ji}_{23} \sin^2\Delta_{23} -
4\mbox{Re}J^{ji}_{31} 
\sin^2\Delta_{31}, \nn \\
\\
\PCPV(\nu_{i}\rightarrow\nu_{j})
& = &-8\sigma_{ij} J \sin\Delta_{12} 
\sin\Delta_{23} \sin\Delta_{31}, \nn
\eea 
with $J$ the Jarlskog invariant and
\begin{equation}
  \label{eq:5}
 J^{ij}_{kh} \equiv U_{ik} U_{kj}^\dagger
U_{jh} U_{hi}^\dagger \, ,\quad
\Delta_{ij}\equiv \Delta m^2_{ij}
L/4E \, ,\quad
\sigma_{ij}\equiv \sum_k \varepsilon_{ijk} 
\end{equation}
 
In order to account for all reported neutrino anomalies one needs only 
two squared mass differences. The LSND result can be understood in
terms the mass difference 
\bea\label{DM}
\Delta M^2 := m_3^2 -m_2^2 \approx 0.3\mbox{ eV}^2
\eea
The second mass difference is tuned so as explain the observed
deficit of electron neutrinos coming from the sun,
\bea\label{dm}
 10^{-4}\mbox{ eV}^2\le\Delta m^2\le 10^{-3}
\eea
with $m_2^2 -m_1^2 \equiv \Delta m^2$, while the atmospheric
neutrino data depend on both differences and are found in agreement
with the prediction. Given the squared mass differences, the mixing angles
are easily found to be \cite{nos} (solution I)
\bea\label{angles}
\theta_{12}\approx 35.5^0\, ,\quad \theta_{23}\approx 27.3^0\, ,\quad
\theta_{13} \approx 13.1^0\; .
\eea
This solution is favoured by the existing oscillation data and, thus, 
implies simultaneous and strong mixing of all three flavours.

\section{Matter effects}

Of all neutrino species, only electron neutrinos can scatter
elastically in the forward direction off electrons in matter, via
charged current interactions. When the electron neutrinos oscillate into
either muon or tau neutrinos, this introduces an additional term in 
the diagonal element of the neutrino flavour evolution matrix 
corresponding to  $\nu_e \rightarrow \nu_e$
\cite{ar}. It is useful to define an effective mass term which stems
from elastic scattering due to charged weak currents,
\bea
a = 2 \sqrt{2}
G_F n_e E =  7.7 \cdot 10^{-5} \mbox{eV}^2 
\left(\frac{\rho}{ \mbox{gr/cm$^3$}}\right) 
\left(\frac{E_\nu}{\mbox{GeV}} \right)
\eea 
where $n_e$ is the electron number density in matter of density 
$\rho$ and $E$ is the neutrino energy.

As is well-known matter effects become important \cite{mat} only when
$a$ is comparable to, or larger than, the
quantity $\Delta_{m_{ij}^2}= m_i^2-m_j^2 $  
for some mass difference and neutrino  energy. Given our 
neutrino mass spectrum and taking into account that for the Earth's 
crust $\rho \simeq  3{ \mbox{gr/cm$^3$}}$, we are far from being
in a range where matter effects would be dominant and could not be
neglected. This would be the case, for example, in the case
of  $\nu_\mu \rightarrow \nu_e$ where large CP asymmetries are
expected but will be masked by matter effects. 
In the case that we discuss here,  matter effects are not
expected to play a significant role but we will include them
anyway. (In the case of electron neutrinos oscillating into tau
neutrinos, and with our parameters, they represent at most a 2~\%
effect.)

In order to exhibit the essential mechanisms we assume for a moment
the two lightest neutrinos to be degenerate in mass. Indeed, this
limiting case is close to the realistic situation where $\Delta M^2
\gg \Delta m^2$, cf. eqs.~(\ref{DM}), (\ref{dm}).  In this case, the 
transition probabilities in vacuum for the ``terrestrial'' experiments
depend only on three variables, i.e., $\theta_{23}$, $\theta_{13}$
and $\Delta M^2 $, as follows,
\bea
P \left( \nu_e \rightarrow \nu_\tau \right) & \approx & 
4U_{13}^2U_{33}^2\;\sin^2 \left( \frac{\Delta M^2 L}{4E}\right)
\nonumber \\
 & = & \cos^2 (\theta_{23}) \;
\sin^2 (2 \theta_{13} ) \; \sin^2 \left( \frac{\Delta M^2 L}{4E} 
\right)\, , \\
P \left( \nu_\mu \rightarrow \nu_\tau \right) & \approx &
4U_{23}^2U_{33}^2\; \sin^2 \left( \frac{\Delta M^2 L}{4E}\right)
\nonumber \\
 & = & \cos^4 (\theta_{13}) \;
\sin^2 (2 \theta_{23} ) \; \sin^2 \left( \frac{\Delta M^2 L}{4E} 
\right)\, . 
\label{osc}
\eea
Note that in this limit of setting $\Delta m^2$ equal to zero the
probabilities are independent of the angle $\theta_{12}$.

When matter effects are included the above formulae are still
valid provided one makes the replacements,
\bea
\Delta M^2 \longrightarrow \Delta M^2 + \frac{\left(3 s_{13}^2 -1 \right) a} 
{2} \\
s_{13}^2 \longrightarrow s_{13}^2 \left[ 1 -  
\frac{2 a \left( s_{13}^2 -1 \right)}{\Delta M^2}\right]\, ,
\eea
with $s_{23}$ unchanged. Here we have assumed that $\Delta M^2 \gg a, \Delta
m^2$, a hierarchy which is respected by the mass spectrum we chose.

From these formulae it is clear that the probability for a muon neutrino
to oscillate into a tau neutrino in matter will not be different
from its vacuum value. The reason for this is clear: the
interpretation of the modulation of neutrino events with zenith angle
reported by SuperKamiokande in terms of a muon neutrino oscillating
into a tau neutrino either requires a nearly maximal $\theta_{23}$, in
two-neutrino mixing schemes, or still a sizeable one, in three-neutrino
mixing schemes. (As noted above, the data strongly disfavour schemes
with three active and one sterile neutrinos.) Furthermore, reactor 
experiments set a strict upper bound on $\theta_{13}$, giving 
$ \sin^2 \theta_{13} \leq .005$. Thus, the factor $\cos^4
(\theta_{13})$ in eq.~(\ref{osc}), to a good approximation, is
equal to 1. Likewise the factor $ \sin^2 (2 \theta_{23} )$ in the same
equation is also close to 1. Therefore, $\nu_\mu \to \nu_\tau$
oscillations are favoured and the influence of matter effects on them
is negligible. 

Introducing the terms containing $\Delta m^2$ does not modify this
simple picture because they are suppressed by the huge gap between
the two squared mass differences. It is important to notice that
although we have assumed degeneracy between the two lightest
neutrinos, this degeneracy is broken by matter effects which introduce 
an ``effective'' mass difference of the order $ a (s_{13}^2 -1)/2 $.

In the case of electron neutrinos oscillating into tau neutrinos, however,
the story is different. In this case, the factor $\sin^2 (2 \theta_{13} )$
in eq.~(10) is a suppressing one and could compensate, at least partially,
the gap in the squared mass differences. With our parameters, the
contributions proportional to $\Delta m^2$ do not really compete with
the one proportional to $\Delta M^2$ but they account for a sizeable
correction. In our results presented below we use the full expression
for the transition probabilities. 

At this point it is important to stress, that unlike our scheme,
in schemes where only two anomalies are taken into account, disregarding
the Los Alamos result, the typical mass differences are $\Delta M^2 \approx
10^{-3} \mbox{ eV}^2$ and  $\Delta m^2 \approx 
10^{-10} \mbox{ eV}^2$, or $10^{-6} \mbox{ eV}^2$ to $10^{-5} \mbox{ eV}^2$, 
depending on whether in the vacuum solution the small-angle
MSW solution or the large-angle MSW solution to the solar neutrino
problem is chosen. Under these assumptions, $\Delta M^2$ is just in
the appropriate range to exhibit sizeable matter effects so that these
cannot be neglected. 

The conclusion so far is clear: If MiniBooNe confirms the LSND evidence
for oscillations, then $\Delta M^2$ is too large to cause a
significant modification of the oscillation probabilities due to
matter effects. As we will see below, the prospects of discovering both 
$\nu_\mu \rightarrow \nu_\tau$ and $\nu_e \rightarrow \nu_\tau$ oscillations
then are promising indeed. However, if KamLAND obtains a positive result in 
disappearance of electron neutrinos, corroborating the large-angle
MSW solution for solar oscillations,  or if KARMEN definitely and
conclusively excludes the LSND result, then $\Delta M^2$ will be in a
range to produce a significant modification of the oscillation
probability due to matter. In this case, the rates for tau appearance
being lower by more than an order of magnitude, an experiment would be 
more difficult and would probably require a more intense neutrino source
than that assumed here.

\section{Event rates}

The calculation of event rates for tau lepton (anti tau lepton)
production from electron neutrinos (anti muon neutrinos) as a result
of oscillations is straightforward. The total number of events in the
two channels is given by 
\bea
n_{\tau^-} = N_{\mu^+} \; N_{kT} \; \frac{10^9 \; N_A \; E_\mu^3}{m_\mu^2 \;
\pi \; L^2} \;
\int_{E_{\mbox{\tiny{min}}}}^{E_\mu}   g_{\nu_e} \; \epsilon \;
\sigma_{\nu_\tau}^{\mbox{\tiny{CC}}} \;  
P\left( \nu_e 
\rightarrow \nu_\tau \right) \; dE
\\
\nn \\
n_{\tau^+} = N_{\mu^+} \; N_{kT} \; \frac{10^9 \; N_A \; E_\mu^3}{m_\mu^2 \;
\pi \; L^2} \;
\int_{E_{\mbox{\tiny{min}}}}^{E_\mu}   g_{\bar{\nu}_\mu} \; \epsilon \; 
\sigma_{\bar{\nu}_\tau}^{\mbox{\tiny{CC}}} \;  
P\left( \bar{\nu}_\mu 
\rightarrow \bar{\nu}_\tau \right) \; dE
\eea
where  $N_{\mu^+}$ is the number of
positive muon decays, $\sigma_{\nu_\tau}$ is the charged current cross
section per nucleon , $ P(\nu_e \rightarrow \nu_\tau)$ and 
$ P\left( \bar{\nu}_\mu  \rightarrow \bar{\nu}_\tau \right)$
are the 
oscillation probabilities for neutrinos traveling inside the Earth taking
into account matter effects. $N_{kT}$ is the size of the detector
in kilotons, $ 10^9 \; N_A $ is the number of nucleons in a kiloton,
$E_\mu$ is the energy of the muons in the ring and  
$E_{\mbox{\tiny{min}}}=$ 5 GeV is a lower cut on the neutrino
energies that we assume, for the sake of an example. 
For our numerical calculation, the energy spectrum (normalized
to 1) of the neutrinos is taken to be
\bea
g_{\nu_e}= 12 x^2 (1-x) \\
g_{\bar{\nu}_\mu} = 2 x^2 (3 -2 x)
\eea
where $x = E_\nu/E_\mu$ is the fractional neutrino energy. It is 
straightforward to obtain the corresponding expression in the
case of either electron anti neutrinos or muon neutrinos.

Regarding the cross section, our calculation is based on the
renormalization group improved parton model, focusing on the inclusive 
process $\nu_\tau (\bar{\nu}_\tau) \; + \; N \longrightarrow \tau^- 
(\tau^+) \; + $ anything. Note that in this case, unlike the charged
current cross section for the muon, the terms proportional to the
charged lepton mass are not negligible.
The differential cross section is \cite{numi},
\bea
\frac{d\sigma}{dx \, dy} &=& \frac{G_F^2 M E_\nu}{\pi} \left( \frac{M_W^2}
{M_W^2 +Q^2}\right)^2 \left\{ F_2 \left( 1 - y - \frac{M x y}{2 E}
\right) + F_1 x y^2 \pm x F_3 \left( y - \frac{y^2}{2} \right) \right. \nn
\\
&&\\
&& 
\left. + \frac{m_{\tau}^2}{M E} \left( - F_2 \left( \frac{M}{4 E} +
\frac{1}{2 x} \right) + \frac{F_1 y}{2} \mp \frac{F_3 y}{4} \right)
\right\} \nn
\eea
where we use the Bjorken scaling variables $x=Q^2/2 M \nu $
and $y=\nu/E_\nu$. Here $-Q^2$ is the invariant momentum transfer between
the incident neutrino and outgoing tau, $\nu =E_\nu - E_\tau $ is the energy
loss in the laboratory frame while $M$ and $M_W$ are the nucleon and 
W-gauge boson masses respectively.

The Callan-Gross relation $ 2 x F_1 = F_2 $ simplifies the above 
equation further. It is then dependent on only two form factors,
$F_2$ and $F_3$, which are given in terms of the parton distributions
by
\bea
F_2 &=& \sum_i x \left( q_i + \bar{q}_i \right) \\
F_3 &=& \sum_i \left( q_i - \bar{q}_i \right)
\eea
For our calculations we have used the MRST99 \cite{MRST} parton distribution.
Although there does not seem to be a general consensus about the best way
of combining the quasi-elastic resonance and the deep inelastic 
scattering cross sections at fixed
energy to form the total cross section, a number of different approaches
were proposed in the literature or were used in practice \cite{32}.
We adopt the simplest of them, and include only the deep inelastic
part. In any case, the error one makes in the prediction due
to uncertainties in the neutrino spectrum and by assuming a constant
matter distribution in the Earth's crust is larger than the uncertainties
in the cross section itself.

\section{Results and conclusions }

We consider a 20 GeV muon storage ring at CERN with a neutrino beam
from one of its straight sections that points to the Gran Sasso
underground laboratory, at a distance of 732 km.  This is about the
same distance as the one between FNAL and the Soudan mine.

In order to mimic a nearly realistic experimental situation
we set a lower cutoff in energy at 5~GeV and we assume a detection
efficiency of 30\%. Also, for two cases, 
production of $\tau^+$ from
the process $\overline{\nu}_\mu\to\overline{\nu}_\tau$ and 
production of $\tau^-$ from
the process $\nu_e\to \nu_\tau$,  
 we show our
results in bins of 3~GeV so as to get a feeling for the number of
events to be expected with our parameters.

Fig.~1 shows the predicted spectrum for tau lepton appearance coming
from $\nu_e \to \nu_\tau$, while Fig.~2 shows the predicted spectrum
of anti taus coming from  $\overline{\nu}_\mu \to \overline{\nu}_\tau$.
For the sake of comparison in Figs.~3 and 4 we plot the same
observables for the choice of parameters reported in Ref~\cite{belen}
which disregards the LSND data (note the different scales in the
ordinates!). Results similar to these were also reported in
\cite{barger}. Fig.~5, finally, shows our result for $\tau^-$
and $\tau^+$
production, grouped in bins of 3~GeV each. 

Provided the difference $\Delta M^2$ is in the range required by the
LSND data, it is clear from Figs.~1, 2, and 5 that an experiment using
a 20~GeV muon storage ring and a baseline of about 730~km between
source and detector (corresponding to the distance CERN-Gran Sasso, or
FNAL to the Soudan mine) would have a fair chance to see the
appearance of charged tau leptons. The average oscillation probability
could be measured with a statistical precision better than 3\%
\cite{juanjo} and $\Delta M^2$ could be determined with a precision of
about 3\%. Note that since matter effects are small in this case, there
is no additional uncertainty on this mass difference arising from
an incomplete knowledge of the oscillation mode.

All in all, with the choice of parameters that we obtained by a
simultaneous explanation of all neutrino anomalies, a 10 kT detector
some 732 km downstream would probe 
$\cos^2 (\theta_{23})$ $\sin^2 (2 \theta_{13} )$ to very low values.
When combined with measurements of $\nu_e \rightarrow \nu_\mu $
oscillations (see for example \cite{lin2}) and with further results
from atmospheric neutrinos, a precise determination of the complete
mass spectrum of neutrinos and of all mixing angles seems possible.

\begin{center}
{\bf Acknowledgements}
\end{center}

We are very grateful to Jos\'e Bernabeu, Daniel de Florian and 
Karl Jakobs for  enlightening
discussions, comments and explanations. 
Financial support from the DFG is also acknowledged.

\newpage

\begin{figure}[!ht]
  \begin{center}
  \epsfig{file=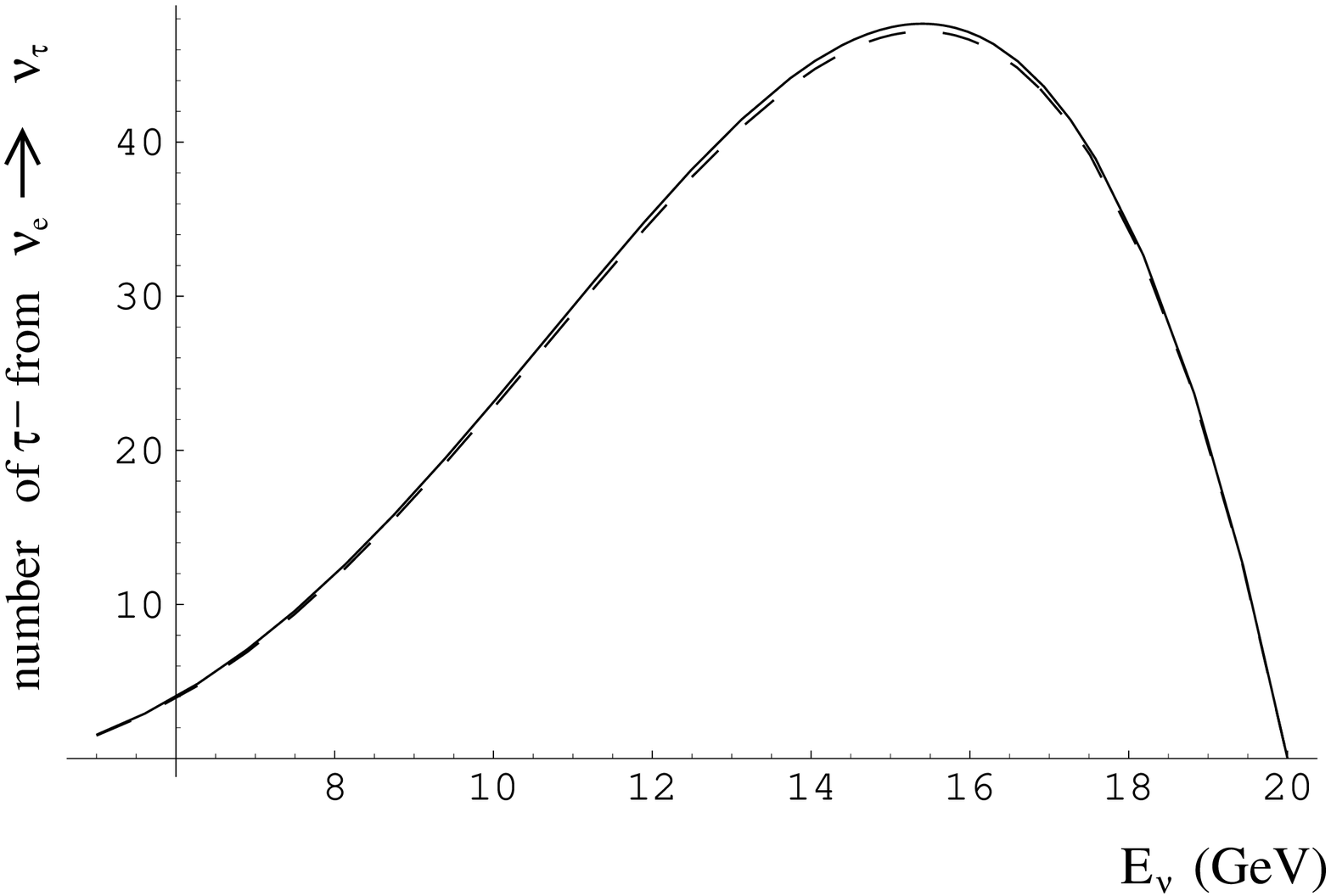,width=9cm}
\caption{Number of tau leptons detected in one year's time
 in a 732 km baseline and assuming 30\% detecting efficiency.  
The solid line correspond to setting 
 the CP violating phase  $\delta=0$
while the dashed line correspond to $\delta=\pi/2$ } 
  \end{center}
\end{figure}

\begin{figure}[!ht]
  \begin{center}
  \epsfig{file=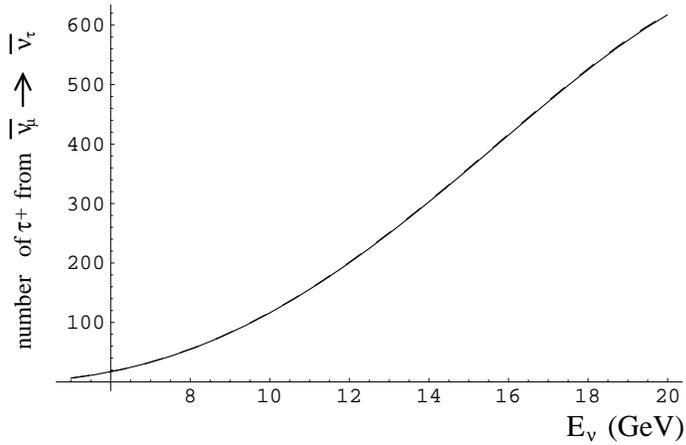,width=9cm}
  \parbox{15cm}{\caption{Number of antitau leptons detected in one year's time
in a 732 km baseline and assuming 30\% detecting efficiency.  
The solid and the dashed curves (defined as in Fig.1) 
are indistinguishable.
}}
  \end{center}
\end{figure}

\pagebreak

\begin{figure}[!ht]
  \begin{center}
  \epsfig{file=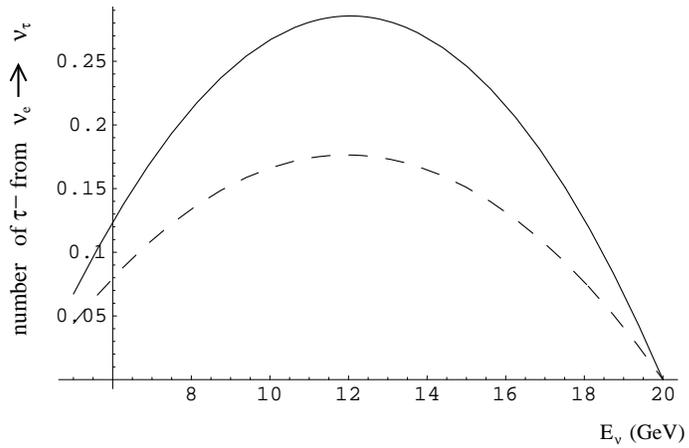,width=9cm}
  \parbox{15cm}{\caption{Number of tau leptons detected in one year's time
 in a 732 km baseline and assuming 30\% detecting efficiency for the
parameters of  ref.~~\cite{belen}. Dashed and solid lines as before.
}}
  \end{center}
\end{figure}

\begin{figure}[!ht]
  \begin{center}
  \epsfig{file=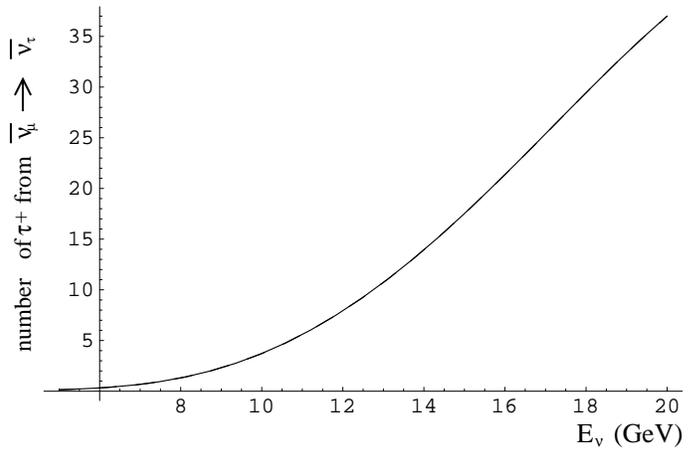,width=9cm}
  \parbox{15cm}{\caption{Number of tau leptons detected in one year's time
 in a 732 km baseline and assuming 30\% detecting efficiency for the
parameters of  ref.~~\cite{belen}.
The solid and the dashed curves are indistinguishable.
}}
  \end{center}
\end{figure}

\begin{figure}[!ht]
  \begin{center}
  \epsfig{file=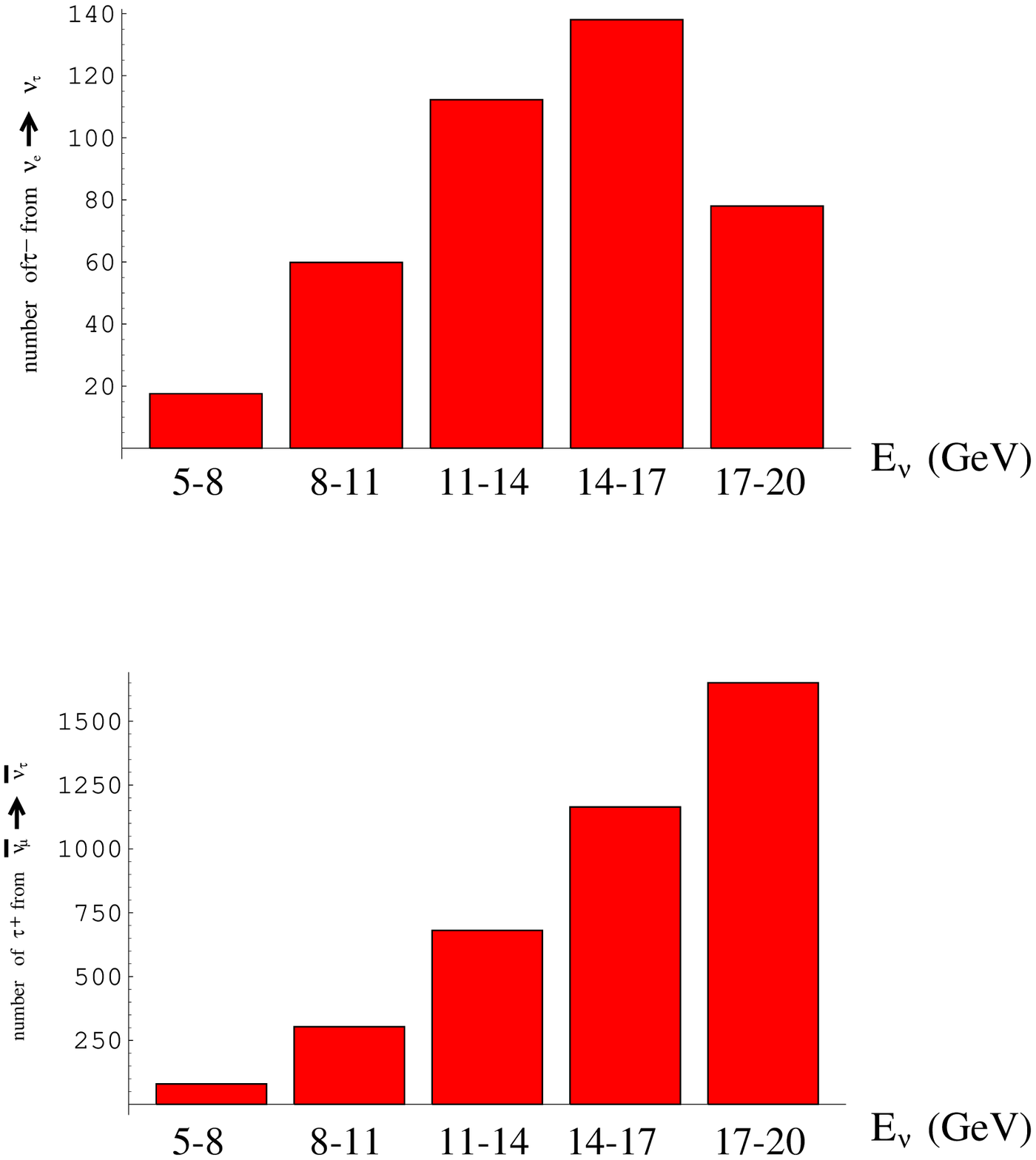,width=9cm}
  \parbox{15cm}{Figure 5: Number of tau (above) and antitau
(below)  leptons  detected in one year's time
 in a 732 km baseline and assuming 30\% detecting efficiency grouped
in bins of 3 GeV.
}
\end{center}
\end{figure}

\end{document}